\documentclass[11pt,a4paper]{article}

\RequirePackage{ifpdf} 
\usepackage{amsmath} 
\usepackage{mathtools}

\usepackage{jheppub}
\usepackage{pstricks}
\usepackage[final]{pdfpages} 
\usepackage{ifpdf} 
\usepackage{slashed}
\usepackage{hyperref}

\usepackage{color} 
\usepackage{graphics}

\usepackage{etoolbox} 
\usepackage{fixmath}

\usepackage{caption} 
\usepackage{subcaption} 
\usepackage{amsfonts}

\usepackage{multirow}
\usepackage{epstopdf}

\usepackage{relsize}
\usepackage{float}

\usepackage{tikz}
\usetikzlibrary{positioning,arrows}
\usetikzlibrary{decorations.pathmorphing}
\usetikzlibrary{decorations.markings}
\usetikzlibrary{shapes.geometric}
\usepackage{endnotes}
\tikzset{
    vector/.style={decorate, decoration={snake}, draw},
    provector/.style={decorate, decoration={snake,amplitude=2.5pt}, draw},
    antivector/.style={decorate, decoration={snake,amplitude=-2.5pt}, draw},
    fermion/.style={draw=black,
      postaction={decorate},decoration={markings,mark=at position .55
        with {\arrow[draw=black]{>}}}}, 
    fermionbar/.style={draw=black, postaction={decorate},
                       decoration={markings,mark=at position .55 with {\arrow[draw=black]{<}}}},
    fermionnoarrow/.style={draw=black},
    gluon/.style={decorate, draw=black,decoration={coil,amplitude=4pt, segment length=6pt}},
    scalar/.style={dashed,draw=black,
      postaction={decorate},decoration={markings,mark=at position .55
        with {\arrow[draw=black]{>}}}}, 
    scalarbar/.style={dashed,draw=black,
      postaction={decorate},decoration={markings,mark=at position .55
        with {\arrow[draw=black]{<}}}}, 
    scalarnoarrow/.style={dashed,draw=black},
    electron/.style={draw=black,
      postaction={decorate},decoration={markings,mark=at position .55
        with {\arrow[draw=black]{>}}}}, 
    bigvector/.style={decorate, decoration={snake,amplitude=4pt}, draw},
}

\title{Two-loop massless QCD corrections to the $g+g \rightarrow H+H$ four-point amplitude}

\author[a]{Pulak Banerjee,}
\author[b]{Sophia Borowka,}
\author[a]{Prasanna K. Dhani,}
\author[c]{Thomas Gehrmann,}
\author[a]{and V. Ravindran}

\affiliation[a]{The Institute of Mathematical Sciences, HBNI, Taramani, Chennai-600113, India}
\affiliation[b]{Theoretical Physics Department, CERN, CH-1211 Geneva, Switzerland}
\affiliation[c]{Physik-Institut, Universit\"at Z\"urich, Winterthurerstrasse 190, CH-8057 Z\"urich, Switzerland  }

\emailAdd{bpulak@imsc.res.in}
\emailAdd{sophia.borowka@cern.ch} 
\emailAdd{prasannakd@imsc.res.in}
\emailAdd{thomas.gehrmann@uzh.ch} 
\emailAdd{ravindra@imsc.res.in}

\abstract{We compute the two-loop massless QCD corrections to 
the four-point amplitude  $g+g \rightarrow H+H$ resulting from effective operator insertions that describe
the interaction of a Higgs boson with gluons in the infinite top quark mass limit. 
This amplitude is an essential ingredient to the third-order QCD corrections to 
Higgs boson pair production.
We have implemented our results in a numerical code that can be used for further phenomenological studies. 
}
\begin{document}

\preprint{\hspace{6 cm} CERN-TH-2018-196, IMSc/2018/09/06,  ZU-TH 32/18}

\allowdisplaybreaks[4]
\unitlength1cm
\maketitle
\flushbottom

\def\D{{\cal D}}
\def\g{\overline {\cal G}}
\def\gm{\gamma}
\def\ep{\epsilon}
\def\zo{\overline{z}_1}
\def\zt{\overline{z}_2}
\def\zob{\overline{z}_1}
\def\ztb{\overline{z}_2}
\def\C{{C}}
\def\C{{C}}
\def\Aob{\overline A_1^I}
\def\Atb{\overline A_2^I}
\def\Athb{\overline A_3^I}
\def\Afb{\overline A_4^I}
\def\Ao{A_1^I}
\def\At{A_2^I}
\def\Ath{A_3^I}
\def\fo{f_1^I}
\def\ft{f_2^I}
\def\fth{f_3^I}
\def\Af{A_4^I}
\def\Bo{B_1^I}
\def\Bt{B_2^I}
\def\Bth{B_3^I}
\def\Dob{\overline D_1^I}
\def\Dobd{\overline D_{d,1}^I}
\def\Dtb{\overline D_2^I}
\def\Dtbd{\overline D_{d,2}^I}
\def\Dthb{\overline D_3^I}
\def\Dthbd{\overline D_{d,3}^I}
\def\btob{\overline \beta_1}
\def\bttb{\overline \beta_2}
\def\btthb{\overline \beta_3}
\def\lfr{\log\left({\mu_F^2 \over \mu_R^2}\right)}
\def\lfrt{\log^2\left({\mu_F^2 \over \mu_R^2}\right)}
\def\lfrtt{\log^3\left({\mu_F^2 \over \mu_R^2}\right)}
\def\lqr{\log\left({q^2 \over \mu_R^2}\right)}
\def\lqrt{\log^2\left({q^2 \over \mu_R^2}\right)}
\def\lqrtt{\log^3\left({q^2 \over \mu_R^2}\right)}
\def\lmt{\log\left({\mu_R^2 \over m_t^2}\right)}
\def\lmts{\log^2\left({\mu_R^2 \over m_t^2}\right)}
\def\lw{\ln(1-\omega)}
\def\w{\omega}
\def\one{\ln(w)}
\def\two{\ln^2(w)}
\def\three{\ln^3(w)}
\def\four{\ln^4(w)}
\def\five{\ln^5(w)}
\def\six{\ln^6(w)}
\def\aLqf{\log\left({q^2 \over \mu_F^2}\right)} 
\def\aLqftwo{\log^2\left({q^2 \over \mu_F^2}\right)} 
\def\aLqfthree{\log^3\left({q^2 \over \mu_F^2}\right)} 
\def\M{{\cal M}}
\def\ep{\epsilon}
\def\unM{\hat{\cal M}}
\def\unas{ \left( \frac{\hat{a}_s}{\mu^{\epsilon}} S_{\epsilon} \right) }
\def\rnM{{\cal M}}
\def\rnas{ \left( a_s  \right) }
\def\b0{\beta_0}
\def\cD{{\cal D}}
\def\cC{{\cal C}}
\def\ca{\text{\tiny C}_\text{\tiny A}}
\def\cf{\text{\tiny C}_\text{\tiny F}}

\def\spt{(s+t)}
\def\spu{(s+u)}
\def\tpu{(t+u)}

\let\footnote=\endnote
\renewcommand*{\thefootnote}{\fnsymbol{footnote}}


\section{Introduction}
The discovery of the Higgs boson at the Large Hadron Collider \cite{Aad:2012tfa,Chatrchyan:2012xdj}
is an important milestone in particle physics.  It 
puts the Standard Model (SM) in a firm position to describe the dynamics of all the known elementary particles. 
Of course, there are several shortcomings in the SM which lead physicists to explore physics beyond the SM.  
There have been tremendous efforts to construct models that address these shortcomings and at the same time 
demonstrate rich phenomenology that can be explored at present and future colliders.  All these culminated into
dedicated experimental searches for hints of new physics which in turn constrain the parameters of beyond the SM scenarios 
\cite{Englert:2014uua}.    

By measuring the mass of the Higgs boson, one can predict the trilinear self-coupling in the Higgs sector of the SM.      
This is a crucial parameter that describes the shape of the Higgs potential.  In order to better understand the Higgs sector and
the nature of the electroweak 
symmetry breaking mechanism, it is important to measure this self-coupling independently.  At hadron colliders,
one of the potential channels that can probe this self-coupling is the production of a pair of Higgs bosons 
\cite{Dawson:1998py,Djouadi:1999gv,Djouadi:1999rca,Muhlleitner:2000jj}.  The dominant production channel
in the SM is through the loop-induced 
gluon fusion subprocess \cite{Glover:1987nx,Plehn:1996wb}.  
At leading order (LO), this process 
involves two mechanisms, with the scattering amplitude for
one of the them depending on  the trilinear Higgs boson coupling.  Since both 
mechanisms are
loop-induced through heavy quarks and there is destructive interference between their respective 
amplitudes, 
the SM production cross section at LHC energies is only few tens of a femtobarn.  In addition, a 
large and irreducible background \cite{Baglio:2012np,Barger:2013jfa,Dolan:2012rv,Papaefstathiou:2012qe,
deLima:2014dta,Behr:2015oqq} makes its detection an experimentally demanding task.  Double Higgs boson production 
can receive substantial contributions from physics processes beyond the SM,  
 and there are already several detailed studies indicating scenarios for a substantial increase 
in its production rate (see \cite{Grober:2017gut} and the references therein).   

Theoretically, it is a challenging task to
compute higher order QCD effects when taking into account the exact top quark mass dependence, since the 
Born-level contribution appears only at one loop. 
The first computation of next-to-leading order (NLO)  QCD corrections was performed  in the
infinite top quark mass limit  in \cite{Dawson:1998py}. In this limit, the top quark is integrated out, resulting in 
a field theory that contains effective operators coupling the Higgs field to the gluon field. 
These early results were then improved upon
by considering various NLO contributions from finite top quark mass effects~\cite{Grigo:2013rya,Frederix:2014hta,Maltoni:2014eza,Degrassi:2016vss,Grober:2017uho,Bonciani:2018omm}. Recently, the full NLO corrections with exact top quark mass dependence could be 
completed~\cite{Borowka:2016ehy, Borowka:2016ypz}, owing to technical progress in the numerical evaluation of 
two-loop integrals and amplitudes with internal masses. 
 At next-to-next-to-leading order (NNLO) level, results are available only 
in the heavy top limit.  The prediction at NNLO level in the soft plus virtual (SV) approximation can be found in  
\cite{deFlorian:2013uza}, the leading top quark mass corrections  were then included in \cite{Grigo:2015dia}, while
in \cite{deFlorian:2013jea} the impact of the remaining hard contributions were studied.  
The relevant Wilson coefficients at NNLO  were obtained in \cite{Grigo:2014jma}.  
For the fully differential results at NNLO level, see \cite{Li:2013flc,Maierhofer:2013sha,deFlorian:2015moa}. 
By using a re-weighting approach, these fixed-order NNLO results for infinite top quark mass can be combined with the exact 
NLO top quark mass dependence to quantify~\cite{Grazzini:2018bsd} the top quark mass effects at NNLO. 
Effects of threshold resummation at next-to-next-to-leading logarithm (NNLL) level using soft collinear effective theory were 
obtained in \cite{deFlorian:2015moa,Shao:2013bz}.

Going beyond NNLO level in QCD is a challenging task owing to the technical difficulties involved in computing
the loop integrals for the virtual subprocesses and the phase space integrals when there are real emissions. 
In this article we make a first step towards computing the third-order correction to the production of a pair of Higgs 
bosons in the gluon initiated channels. In particular we compute 
virtual amplitudes for the subprocess $g+g \rightarrow H+H$, resulting from two effective operator insertions, 
at the two-loop level. The paper is structured as follows. In Section~\ref{sec:framework}, we introduce 
the notation, describe the effective field theory that results in the limit of an infinite top quark mass, and discuss 
the different purely virtual contributions to Higgs boson pair production up to next-to-next-to-next-to-leading order (N$^3$LO). Section~\ref{sec:calc} 
describes in detail the calculation of the two-loop amplitude for $g+g \rightarrow H+H$, and the numerical 
evaluation of the results is discussed in~\ref{sec:num}.
We conclude with an outlook on future applications in Section~\ref{sec:conc}.


\section{Virtual Higgs Pair Production Contributions to N$^3$LO}
\label{sec:framework}
\subsection{Higgs effective field theory}
We compute the relevant amplitudes in an effective theory where the top quark degrees of freedom
are integrated out. The effective Lagrangian that describes the coupling of one and two 
Higgs bosons to gluons is given by
\begin{eqnarray}
	{\cal L}_{eff} = -{1 \over 4} \left( C_H(a_s) {\phi \over v} - C_{HH}(a_s) {\phi^2 \over v^2}\right)G_{\mu \nu} G^{\mu \nu} \,,
	\label{eq:eff}
\end{eqnarray}
where  $G_{\mu \nu}$ denotes the gluon field strength tensor, $\phi$, the Higgs boson and $v=246$~GeV is the vacuum expectation value of the Higgs field.  Note that we have taken only those terms in the  
${\cal L}_{eff}$ into account that are relevant for the production of two Higgs bosons in a gluon-gluon initiated process.
The constants $C_H$ and $C_{HH}$ are the Wilson coefficients \cite{Grigo:2014jma,Djouadi:1991tka,
Kramer:1996iq,Chetyrkin:1997iv,Spira:2016zna,Gerlach:2018hen} determined by matching the effective theory to the full theory and 
they can be expanded   
in powers of the renormalized strong coupling constant $a_s=g_s^2(\mu_R^2)/(16\pi^2)=\alpha_s(\mu_R^2)/(4\pi)$ with $\mu_R$ the renormalisation scale,
\begin{eqnarray}
	C_H(a_s) &=& -\frac{4 a_s}{3}  \Bigg[ 1 + a_s   \Big( 11\Big)
	\nonumber \\ &&+ a_s^2 \Bigg(  \Bigg\{
           \frac{2777}{18}
          + 19 \lmt
	  \Bigg\}
	+  n_f   \Bigg\{
          - \frac{67}{6}
          + \frac{16}{3} \lmt
	  \Bigg\}
          \Bigg)
	  \nonumber \\
	&&+ a_s^3  \Bigg( 
          - \frac{2892659}{648}
          + \frac{3466}{9} \lmt
          + 209 \lmts
          + \frac{897943}{144} \zeta_3
	  \nonumber \\
	&&	+ n_f   \Bigg\{
           \frac{40291}{324}
          + \frac{1760}{27} \lmt
          + 46 \lmts
          - \frac{110779}{216} \zeta_3
	  \Bigg\}
	  \nonumber\\
	&& + n_f^2   \Bigg\{
          - \frac{6865}{486}
          + \frac{77}{27} \lmt
          - \frac{32}{9} \lmts
	  \Bigg\}
       \Bigg)\Bigg] \,,
\\
	C_{HH}(a_s) &=& -\frac{4 a_s}{3}  \Bigg[ 1+ a_s   \Big( 11
          \Big)\nonumber \\ &&
       + a_s^2 \Bigg(  
           \frac{3197}{18}
          + 19 \lmt
	+  n_f   \Bigg\{
          - \frac{1}{2}
          + \frac{16}{3} \lmt
	  \Bigg\}
        \Bigg)\Bigg]\,,
\end{eqnarray}
and where $n_f$ is the number of light flavors, $m_t$ is the $\overline{MS}$ top quark mass at scale $\mu_R$ and $N=3$ is fixed for QCD.

\subsection{Kinematics}
Consider the production of a pair of Higgs bosons in the gluon fusion subprocess,
\begin{eqnarray}
\label{eq:amp}
	g(p_1)+g(p_2) \rightarrow H(p_3)+H(p_4)\,,
\end{eqnarray}
where $p_1$ and $p_2$ are the momenta of the incoming gluons, and $p_3$ and $p_4$ 
the momenta for the outgoing Higgs bosons, respectively.
The Mandelstam variables for the above process are given by
\begin{eqnarray}
s = (p_1+p_2)^2, \quad t = (p_1-p_3)^2, \quad u=(p_2-p_3)^2\,.
\end{eqnarray}
They satisfy $s+t+u=2 m_h^2$ where $m_h$ is the mass of the Higgs boson.
In the following, we describe the computation of the one- and two-loop 
QCD corrections to the amplitude given in Eq.~(\ref{eq:amp}). 
We find that it is convenient to express this amplitude in terms of the
dimensionless variables $x$, $y$ and $z$
\begin{eqnarray}
	s = m_h^2 {(1+x)^2 \over x},\quad t = -m_h^2 y,\quad u= -m_h^2 z.
\end{eqnarray}

\subsection{Tensors and projectors}
Using gauge invariance, the amplitude can be decomposed in terms of two second rank Lorentz 
tensors ${\cal T}_i^{\mu \nu}$ with $i=1,2$ as follows
\cite{Glover:1987nx}:
\begin{eqnarray}
	{\cal M}^{\mu \nu}_{ab} = \delta_{ab} \left({\cal T}_1^{\mu\nu}~{\cal M}_1  
			    + {\cal T}_2^{\mu\nu}~{\cal M}_2\right)\,, 
\end{eqnarray}
where the tensors are given by
\begin{eqnarray}
	{\cal T}_1^{\mu\nu} & = & g^{\mu \nu} - {1 \over p_1\cdot p_2} \Big(
	p_1^\nu p_2^\mu \Big)
	\\
	{\cal T}_2^{\mu\nu} & = & g^{\mu \nu} + {1 \over p_1\cdot p_2~ p_T^2} \Big(
	m_h^2~ p_2^\mu p_1^\nu - 2 p_1 \cdot p_3~ p_2^\mu p_3^\nu -2 p_2\cdot p_3~ p_3^\mu p_1^\nu
	+2 p_1\cdot p_2~ p_3^\mu p_3^\nu \Big)\,,
\end{eqnarray}
with $p_T^2 = (t u - m_h^4)/s$.  In color space, the  amplitude is diagonal in the indices ($a,b$) of the 
incoming gluons. The scalar functions ${\cal M}_i$ can be obtained from ${\cal M}^{\mu \nu}_{ab}$ by using appropriate
projectors as follows
\begin{eqnarray}
	{\cal M}_i={1 \over N^2-1} ~P_i^{\mu \nu}{\cal M}^{ab}_{\mu \nu} 
	\delta_{ab},\quad \quad \quad i=1,2 \,,
\end{eqnarray}
where the projectors in $d$ dimensions are given by,
\begin{eqnarray}
	P_1^{\mu \nu} &=& \frac{1}{4}\frac{d-2}{d-3} {\cal T}_1^{\mu\nu} -\frac{1}{4}\frac{d-4}{d-3}{\cal T}_2^{\mu\nu}\,,
	\nonumber \\
	P_2^{\mu \nu} &=& -\frac{1}{4}\frac{d-4}{d-3}{\cal T}_1^{\mu\nu}+\frac{1}{4}\frac{d-2}{d-3}{\cal T}_2^{\mu\nu}\,.
\label{eq:proj}
\end{eqnarray}

\subsection{Diagrams to $\mathcal{O}(a_s^4)$}

When considering higher order massless QCD corrections to the $g+g \to H+H$ amplitudes in the effective theory, 
we  encounter 
two topologically distinct classes of subprocesses we call Class-A 
and Class-B hereafter.  We perform  an expansion in $a_s$ to include all the contributing diagrams.


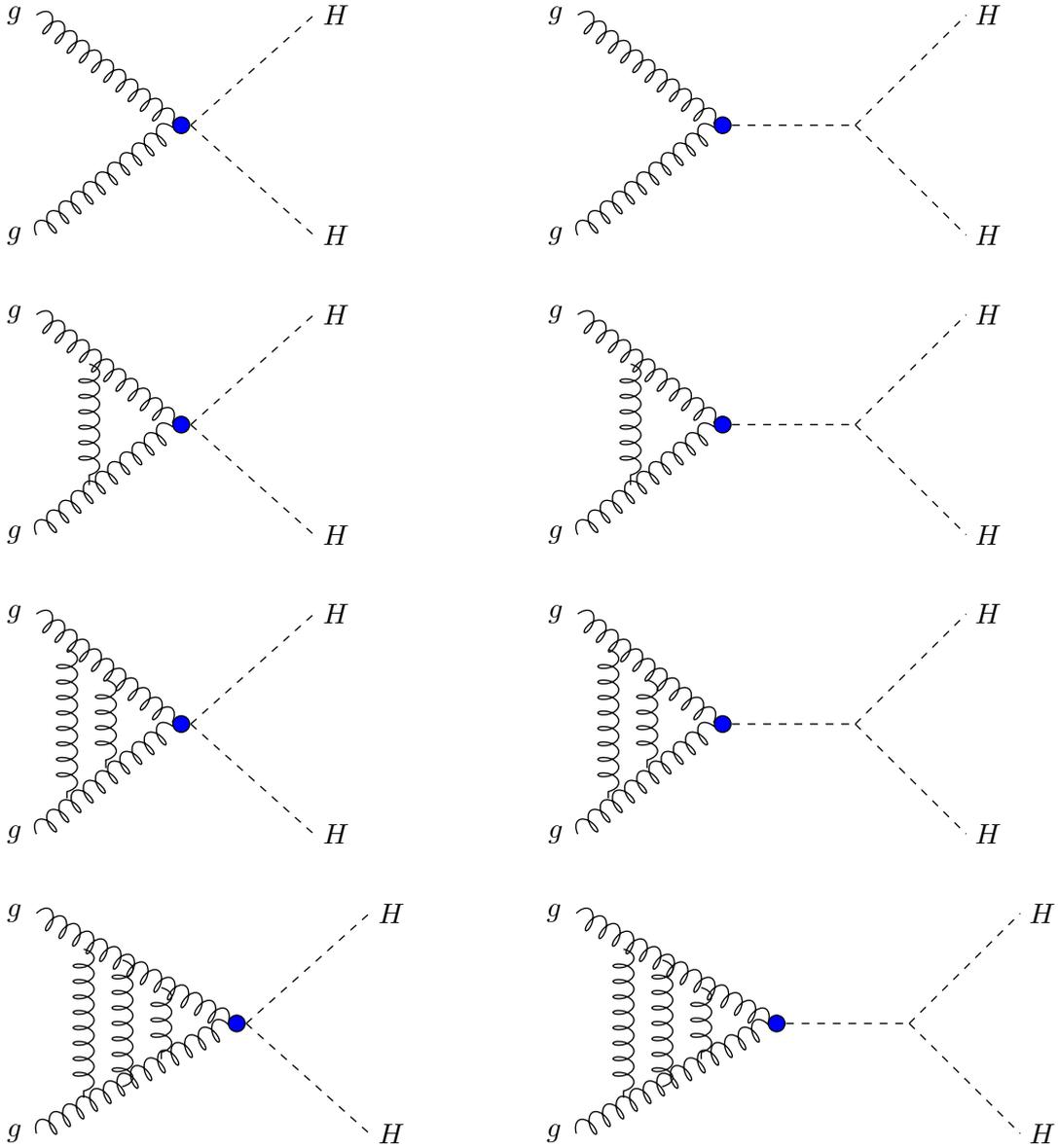
\begin{figure}[htb!]
\begin{centering}
\begin{tikzpicture}[line width=0.5 pt, scale=0.75]
\draw[gluon] (-2.5,2.0) -- (0,0);
\draw[gluon] (-2.5,-2.0) -- (0,0);
\draw[fill=blue] (0.11,0) circle (.15cm);
\draw[scalarnoarrow] (0.28,0)--(2.5,2.0);
\draw[scalarnoarrow] (0.28,0)--(2.5,-2.0);
\node at (-2.9,2.0) {$g$};
\node at (-2.9,-2.0) {$g$};
\node at (2.9,2.0) {$H$};
\node at (2.9,-2.0) {$H$};
 \end{tikzpicture}
\quad \quad \qquad \qquad 
\begin{tikzpicture}[line width=0.5 pt, scale=0.75]
\draw[gluon] (-2.5,2.0) -- (0,0);
\draw[gluon] (-2.5,-2.0) -- (0,0);
\draw[fill=blue] (0.11,0) circle (.15cm);
\draw[scalarnoarrow] (0.28,0)--(2.5,0);
\draw[scalarnoarrow] (2.5,0)--(4.5,2.0);
\draw[scalarnoarrow] (2.5,0)--(4.5,-2.0);
\node at (-2.9,2.0) {$g$};
\node at (-2.9,-2.0) {$g$};
\node at (4.9,2.0) {$H$};
\node at (4.9,-2.0) {$H$};
 \end{tikzpicture}
 \end{centering}
%
%
\\
\\
%
%
\begin{centering}
\begin{tikzpicture}[line width=0.5 pt, scale=0.75]
\draw[gluon] (-2.5,2.0) -- (0,0);
\draw[gluon] (-2.5,-2.0) -- (0,0);
\draw[gluon] (-1.55,1.1) -- (-1.55,-1.1);
\draw[fill=blue] (0.11,0) circle (.15cm);
\draw[scalarnoarrow] (0.28,0)--(2.5,2.0);
\draw[scalarnoarrow] (0.28,0)--(2.5,-2.0);
\node at (-2.9,2.0) {$g$};
\node at (-2.9,-2.0) {$g$};
\node at (2.9,2.0) {$H$};
\node at (2.9,-2.0) {$H$};
 \end{tikzpicture}
\quad \quad \qquad \qquad 
\begin{tikzpicture}[line width=0.5 pt, scale=0.75]
\draw[gluon] (-2.5,2.0) -- (0,0);
\draw[gluon] (-2.5,-2.0) -- (0,0);
\draw[gluon] (-1.55,1.1) -- (-1.55,-1.1);
\draw[fill=blue] (0.11,0) circle (.15cm);
\draw[scalarnoarrow] (0.28,0)--(2.5,0);
\draw[scalarnoarrow] (2.5,0)--(4.5,2.0);
\draw[scalarnoarrow] (2.5,0)--(4.5,-2.0);
\node at (-2.9,2.0) {$g$};
\node at (-2.9,-2.0) {$g$};
\node at (4.9,2.0) {$H$};
\node at (4.9,-2.0) {$H$};
 \end{tikzpicture}
 \end{centering}
%
\\
\\
%
\begin{centering}
\begin{tikzpicture}[line width=0.5 pt, scale=0.75]
\draw[gluon] (-2.5,2.0) -- (0,0);
\draw[gluon] (-2.5,-2.0) -- (0,0);
\draw[gluon] (-1.25,0.80) -- (-1.25,-0.80);
\draw[gluon] (-1.95,1.35) -- (-1.95,-1.35);
\draw[fill=blue] (0.11,0) circle (.15cm);
\draw[scalarnoarrow] (0.28,0)--(2.5,2.0);
\draw[scalarnoarrow] (0.28,0)--(2.5,-2.0);
\node at (-2.9,2.0) {$g$};
\node at (-2.9,-2.0) {$g$};
\node at (2.9,2.0) {$H$};
\node at (2.9,-2.0) {$H$};
 \end{tikzpicture}
\quad \quad  \qquad \qquad 
\begin{tikzpicture}[line width=0.5 pt, scale=0.75]
\draw[gluon] (-2.5,2.0) -- (0,0);
\draw[gluon] (-2.5,-2.0) -- (0,0);
\draw[gluon] (-1.25,0.80) -- (-1.25,-0.80);
\draw[gluon] (-1.95,1.35) -- (-1.95,-1.35);
\draw[fill=blue] (0.11,0) circle (.15cm);
\draw[scalarnoarrow] (0.28,0)--(2.5,0);
\draw[scalarnoarrow] (2.5,0)--(4.5,2.0);
\draw[scalarnoarrow] (2.5,0)--(4.5,-2.0);
\node at (-2.9,2.0) {$g$};
\node at (-2.9,-2.0) {$g$};
\node at (4.9,2.0) {$H$};
\node at (4.9,-2.0) {$H$};
 \end{tikzpicture}
 \end{centering}
%
%
\\
\\
%
\begin{centering}
\begin{tikzpicture}[line width=0.5 pt, scale=0.75]
\draw[gluon] (-3.5,2.0) -- (0,0);
\draw[gluon] (-3.5,-2.0) -- (0,0);
\draw[gluon] (-1.25,0.65) -- (-1.25,-0.65);
\draw[gluon] (-1.95,1.15) -- (-1.95,-1.15);
\draw[gluon] (-2.65,1.35) -- (-2.65,-1.35);
\draw[fill=blue] (0.11,0) circle (.15cm);
\draw[scalarnoarrow] (0.28,0)--(2.5,2.0);
\draw[scalarnoarrow] (0.28,0)--(2.5,-2.0);
\node at (-3.9,2.0) {$g$};
\node at (-3.9,-2.0) {$g$};
\node at (2.9,2.0) {$H$};
\node at (2.9,-2.0) {$H$};
 \end{tikzpicture}
\quad \quad \qquad
\begin{tikzpicture}[line width=0.5 pt, scale=0.75]
\draw[gluon] (-3.5,2.0) -- (0,0);
\draw[gluon] (-3.5,-2.0) -- (0,0);
\draw[gluon] (-1.25,0.65) -- (-1.25,-0.65);
\draw[gluon] (-1.95,1.15) -- (-1.95,-1.15);
\draw[gluon] (-2.65,1.35) -- (-2.65,-1.35);
\draw[fill=blue] (0.11,0) circle (.15cm);
\draw[scalarnoarrow] (0.28,0)--(2.5,0);
\draw[scalarnoarrow] (2.5,0)--(4.5,2.0);
\draw[scalarnoarrow] (2.5,0)--(4.5,-2.0);
\node at (-3.9,2.0) {$g$};
\node at (-3.9,-2.0) {$g$};
\node at (4.9,2.0) {$H$};
\node at (4.9,-2.0) {$H$};
\end{tikzpicture}
\end{centering}
\caption{Class-A: Tree, one-, two- and three-loop amplitudes}
\label{fig:classa}
\end{figure}
%


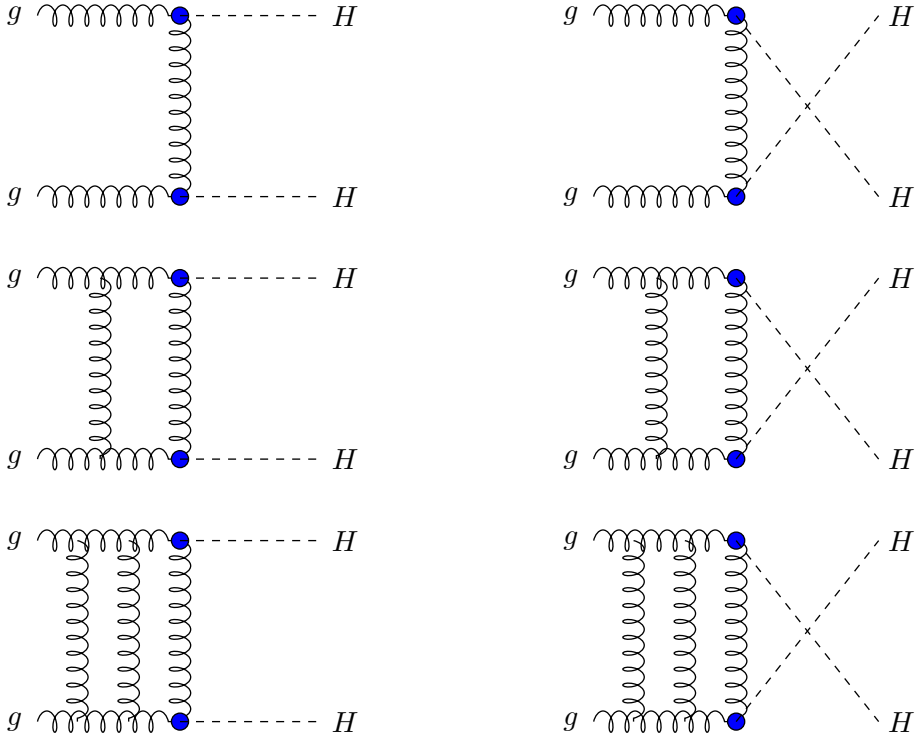
\begin{figure}[htb!]
\begin{centering}
\begin{tikzpicture}[line width=0.5 pt, scale=0.75]
\draw[gluon] (-2.5,1.6) -- (0,1.6);
\draw[gluon] (-2.5,-1.6) -- (0,-1.6);
\draw[gluon] (0,1.6) -- (0,-1.6);
\draw[fill=blue] (0,1.6) circle (.15cm);
\draw[fill=blue] (0,-1.6) circle (.15cm);
\draw[scalarnoarrow] (0,1.6)--(2.5,1.6);
\draw[scalarnoarrow] (0,-1.6)--(2.5,-1.6);
\node at (-2.9,1.6) {$g$};
\node at (-2.9,-1.6) {$g$};
\node at (2.9,1.6) {$H$};
\node at (2.9,-1.6) {$H$};
 \end{tikzpicture}
\quad \quad \qquad \qquad 
\begin{tikzpicture}[line width=0.5 pt, scale=0.75]
\draw[gluon] (-2.5,1.6) -- (0,1.6);
\draw[gluon] (-2.5,-1.6) -- (0,-1.6);
\draw[gluon] (0,1.6) -- (0,-1.6);
\draw[fill=blue] (0,1.6) circle (.15cm);
\draw[fill=blue] (0,-1.6) circle (.15cm);
\draw[scalarnoarrow] (0,1.6)--(2.5,-1.6);
\draw[scalarnoarrow] (0,-1.6)--(2.5,1.6);
\node at (-2.9,1.6) {$g$};
\node at (-2.9,-1.6) {$g$};
\node at (2.9,1.6) {$H$};
\node at (2.9,-1.6) {$H$};
\end{tikzpicture}
\end{centering}
%
\\
\\
%
\begin{centering}
\begin{tikzpicture}[line width=0.5 pt, scale=0.75]
\draw[gluon] (-2.5,1.6) -- (0,1.6);
\draw[gluon] (-2.5,-1.6) -- (0,-1.6);
\draw[gluon] (0,1.6) -- (0,-1.6);
\draw[gluon] (-1.40,1.6) -- (-1.40,-1.6);
\draw[fill=blue] (0,1.6) circle (.15cm);
\draw[fill=blue] (0,-1.6) circle (.15cm);
\draw[scalarnoarrow] (0,1.6)--(2.5,1.6);
\draw[scalarnoarrow] (0,-1.6)--(2.5,-1.6);
\node at (-2.9,1.6) {$g$};
\node at (-2.9,-1.6) {$g$};
\node at (2.9,1.6) {$H$};
\node at (2.9,-1.6) {$H$};
 \end{tikzpicture}
\quad \quad  \qquad \qquad 
\begin{tikzpicture}[line width=0.5 pt, scale=0.75]
\draw[gluon] (-2.5,1.6) -- (0,1.6);
\draw[gluon] (-2.5,-1.6) -- (0,-1.6);
\draw[gluon] (0,1.6) -- (0,-1.6);
\draw[gluon] (-1.40,1.6) -- (-1.40,-1.6);
\draw[fill=blue] (0,1.6) circle (.15cm);
\draw[fill=blue] (0,-1.6) circle (.15cm);
\draw[scalarnoarrow] (0,1.6)--(2.5,-1.6);
\draw[scalarnoarrow] (0,-1.6)--(2.5,1.6);
\node at (-2.9,1.6) {$g$};
\node at (-2.9,-1.6) {$g$};
\node at (2.9,1.6) {$H$};
\node at (2.9,-1.6) {$H$};
\end{tikzpicture}
\end{centering}
\\
\\
%
%
\begin{centering}
\begin{tikzpicture}[line width=0.5 pt, scale=0.75]
\draw[gluon] (-2.5,1.6) -- (0,1.6);
\draw[gluon] (-2.5,-1.6) -- (0,-1.6);
\draw[gluon] (0,1.6) -- (0,-1.6);
\draw[gluon] (-0.9,1.6) -- (-0.9,-1.6);
\draw[gluon] (-1.8,1.6) -- (-1.8,-1.6);
\draw[fill=blue] (0,1.6) circle (.15cm);
\draw[fill=blue] (0,-1.6) circle (.15cm);
\draw[scalarnoarrow] (0,1.6)--(2.5,1.6);
\draw[scalarnoarrow] (0,-1.6)--(2.5,-1.6);
\node at (-2.9,1.6) {$g$};
\node at (-2.9,-1.6) {$g$};
\node at (2.9,1.6) {$H$};
\node at (2.9,-1.6) {$H$};
 \end{tikzpicture}
\quad \quad \qquad \qquad 
\begin{tikzpicture}[line width=0.5 pt, scale=0.75]
\draw[gluon] (-2.5,1.6) -- (0,1.6);
\draw[gluon] (-2.5,-1.6) -- (0,-1.6);
\draw[gluon] (0,1.6) -- (0,-1.6);
\draw[gluon] (-0.9,1.6) -- (-0.9,-1.6);
\draw[gluon] (-1.8,1.6) -- (-1.8,-1.6);
\draw[fill=blue] (0,1.6) circle (.15cm);
\draw[fill=blue] (0,-1.6) circle (.15cm);
\draw[scalarnoarrow] (0,1.6)--(2.5,-1.6);
\draw[scalarnoarrow] (0,-1.6)--(2.5,1.6);
\node at (-2.9,1.6) {$g$};
\node at (-2.9,-1.6) {$g$};
\node at (2.9,1.6) {$H$};
\node at (2.9,-1.6) {$H$};
\end{tikzpicture}
\end{centering}
\caption{Class-B: Tree, one- and two-loop amplitudes}
\label{fig:classb}
\end{figure}

\begin{itemize}
\item Class-A, see Fig.~\ref{fig:classa}, contains diagrams where two  
Higgs bosons couple to each other and to gluons. They either couple to the 
gluons directly through a $C_{HH}$ Wilson coefficient (left-hand column of Fig.~\ref{fig:classa}), or 
through a Higgs boson propagator and the $C_{H}$ Wilson 
coefficient (right-hand column of Fig.~\ref{fig:classa}). The latter diagrams are linearly proportional to the 
triple Higgs coupling $\lambda$. 
\item Class-B, see Fig.~\ref{fig:classb}, contains diagrams where Higgs bosons 
couple to two gluons through the effective vertices proportional to 
$C_H$, but do not couple to each other. 
\end{itemize}
Both Wilson coefficients $C_H$ and $C_{HH}$ start at order $a_s$. Consequently, 
to LO in $a_s$ only Class-A diagrams contribute. 
Beyond LO, that is from order $a_s^2$ onwards, the class-A diagrams are 
only of form factor type and the
results for class-A to $a_s^4$ can be readily obtained from the three loop form 
factor~\cite{Baikov:2009bg,Gehrmann:2010ue} that appears in purely virtual 
contributions to single Higgs boson production. 
The class-B diagrams start contributing from 
order $a_s^2$, with results only available up to order $a_s^3$~\cite{Dawson:1998py}. 
In the following, we will complete the $a_s^4$ contributions to the $g+g \to H+H$ amplitude, by 
computing the class-B diagrams to this order, which amount to their two-loop corrections. 

In general, the scalar amplitudes ${\cal M}_i$ can be written as a sum of amplitudes resulting from the two classes A and B
\begin{eqnarray}
	{\cal M}_i = {\cal M}_i^A + {\cal M}_i^B,\quad \quad \quad i=1,2\,.
\end{eqnarray}
Since the ${\cal M}_i^A$ are proportional to the Higgs boson form factor, they can be 
expressed as
\begin{eqnarray}
	{\cal M}_i^A = \delta_{i1}{\overline {\cal M}_1^{A}}(a_s) \sum_{j=0}^\infty a_s^j {\cal F}^{(j)}(d)\,,
\end{eqnarray}
where
\begin{eqnarray}
	{\overline {\cal M}_1^{A}}(a_s) =i {s \over 2}  \left(C_{HH}(a_s) - C_H(a_s){6 \lambda v^2 \over s-m_h^2} \right)\,. 
\end{eqnarray}
The amplitude ${\cal M}_2^{A}$ is
identically to zero to all orders in perturbation theory 
due to the choice of the tensorial basis.  
The form factors ${\cal F}^{(j)}(d)$ for $j=1,2,3$ are known in the literature
\cite{Baikov:2009bg,Gehrmann:2010ue}.

In this article, the amplitudes of class-B are presented up to two loop level in perturbative QCD.
At each order,
the amplitude contains a pair of vertices resulting from the first term of the effective Lagrangian ${\cal L}_{eff}$ and hence
will be proportional to the square of the Wilson coefficient $C_H(a_s)$, expanded to the desired accuracy in $a_s$.  
Beyond leading order, the one- and two-loop diagrams are not only ultraviolet (UV) divergent but also infrared (IR) 
divergent resulting from 
soft and collinear regions of the loop momenta. We use dimensional regularization to treat 
both UV and IR divergences and all the
divergences show up as poles in $\epsilon$, where the space time dimension is $d=4-2\epsilon$. 

\subsection{Ultraviolet renormalization and operator mixing}
\label{sec:ren}

The bare strong coupling constant in the 
regularized theory is denoted by $\hat a_s$ which is related to its renormalized counter-part by
%
%
\begin{align} \label{renas}
 \hat{a}_s\mu^{2\epsilon} S_{\epsilon} &= a_s\mu_R^{2\epsilon} Z(\mu_R^2)
\nonumber\\[1ex]
 &= a_s\mu_R^{2\epsilon} \left[   1 - a_s \left( \frac{\b0}{\ep} \right) 
  + a_s^2 \left( \frac{\b0^2}{\ep^2} - \frac{\beta_1}{2\ep} \right) + {\cal O}(a_s^3)  \right]\,, 
\end{align}
%
where $S_{\epsilon} = {\rm exp} \left[(\ln 4\pi-\gamma)\epsilon
\right]$ with $\gamma \approx 0.5772...$ the Euler-Mascheroni constant. 
The beta function coefficients $\beta_0$ and $\beta_1$ are given
by
\begin{align}
\beta_0&={11 \over 3 } C_A - {4 \over 3 } T_F n_f \, ,
\nonumber \\[0.5ex]
\beta_1&={34 \over 3 } C_A^2-4 T_F n_f C_F -{20 \over 3} T_F n_f C_A \, ,
\end{align}
for the SU(N) color factors we have
\begin{eqnarray}
	C_A = N,\quad \quad C_F = {N^2-1\over 2 N},\quad  {\rm and}  \quad T_F = {1 \over 2} \,.
\end{eqnarray}
Besides coupling constant renormalisation, the amplitudes also require the renormalisation of the effective operators 
in the effective Lagrangian, Eq.~(\ref{eq:eff}). 
Both composite operators that appear in our one- and two-loop amplitudes 
can develop UV divergences and thus have to undergo renormalisation, as derived in detail in \cite{Zoller:2016iam}.
In particular, a
new renormalisation constant $Z^L_{11}$ is needed in a counter term proportional to $G_{\mu \nu} G^{\mu \nu}\phi \phi$ 
to renormalize the additional UV divergence resulting from amplitudes involving two
$G_{\mu \nu} G^{\mu \nu} \phi$ type operators starting from 2-loop order in class-B amplitudes.
If we denote the amplitudes computed in the bare theory by $ \hat {\cal M}^{B}_i$, then
the relation between these bare amplitudes and the UV renormalized ones is given by   
\begin{eqnarray}
	{\cal M}^{B}_i = Z^2_{{\cal O}} \hat {\cal M}^{B}_i + Z_{11}^L \hat {\cal M}^{A,(0)}_i {\Big |}_{\lambda=0} \,,
\end{eqnarray}
where $\unM^{A(0)}_i$ is the Born amplitude from class-A and $\unM^{B}_{i}$ are the unrenormalized 
amplitudes from class-B. The latter can be expanded in powers of the unrenormalized coupling constant $\hat a_s$ as
\begin{equation} \label{unm}
	\hat {\cal M}_i^B = \unM_i^{B,(0)} + \Big(\hat{a}_s\mu^{2\epsilon}S_{\epsilon}\Big) \unM_i^{B,(1)}  
	+ \Big(\hat{a}_s\mu^{2\epsilon}S_{\epsilon}\Big)^2 \unM_i^{B,(2)} + {\cal O}(\hat{a}_s^3)\,.  
\end{equation}
The overall renormalisation constant \cite{Nielsen:1975ph,Spiridonov:1988md,Kataev:1981gr} is given by
\begin{align}\label{renl}
	Z_{{\cal O}}&= 1 - a_s \left( \frac{1}{\ep} r_{{\cal O}_{1;1}} \right)
+ a_s^2 \left( \frac{1}{\ep^2} r_{{\cal O}_{2;2}}
     - \frac{1}{\ep} r_{{\cal  O}_{2;1}} \right) 
 + {\cal O}(a_s^3)  \; ,
\end{align}
where 
\begin{equation*}
	r_{{\cal O}_{1;1}} =  \b0 \;, \quad
	r_{{\cal O}_{2;2}} = \b0^2 \;, \quad
	r_{{\cal O}_{2;1}} =  \beta_1 \;, \quad
\end{equation*}
and $Z_{11}^L$ is given by \cite{Zoller:2016iam},
\begin{eqnarray}
	Z_{11}^L =a_s^2 {\beta_1 \over \epsilon} + {\cal O}(a_s^3)\,.
\end{eqnarray}
The UV renormalized amplitude ${\cal M}^{B}_i$ can be expanded in powers of $a_s$ up to the two-loop level as follows:
\begin{equation} 
	\label{renamp}
	\rnM^B_i =  \rnM_i^{B,(0)}+ {a_s} \rnM^{B,(1)}_i+ 
	{a_s^{2}} \rnM^{B,(2)}_i  + {\cal O}({a}_s^3) \,,
\end{equation}
where,
\begin{eqnarray} \label{rln}
	\rnM_i^{B,(0)} &=&  \unM_i^{B,(0)}  \; ,
\nonumber\\
	\rnM_i^{B,(1)} &= & \mu_R^{2\epsilon} \Bigg[
	\unM_i^{B,(1)}-  \frac{1}{\mu_R^{2\epsilon}} \left({1 \over \ep} 2 r_{{\cal O}_{1;1}} \right) \unM_i^{B,(0)}~  \Bigg]\; ,
\nonumber\\
	\rnM_i^{B,(2)}&= &  \mu_R^{4\epsilon} \Bigg[
\unM_i^{B,(2)}  
	- \frac{1}{\mu_R^{2\epsilon}}\left({1 \over \ep} \left(2 r_{{\cal O}_{1;1}} + \beta_0\right) \right) \unM_i^{B,(1)}  
\nonumber\\
	&+&  \frac{1}{\mu_R^{4 \epsilon}}\left(  
	 {1 \over \ep^2} \left(r^2_{{\cal O}_{1;1}}
	 +2 r_{{\cal O}_{2;2}}\right)
	 -{1 \over \ep} \left(2 r_{{\cal O}_{2;1}}
	 \right)
	  \right) \unM_i^{B,(0)}  
\nonumber\\
&+& \frac{1}{\mu_R^{4 \epsilon}} \left({\beta_1 \over \ep} \right) \hat {\cal M}^{A,(0)}_i {\Big |}_{\lambda=0} \Bigg]\,.
\end{eqnarray}
In summary, the UV divergences that appear at the one- and two-loop level 
can be removed using coupling constant renormalisation through $Z$ 
and the overall operator and the contact 
renormalisation constants, $Z_{\cal O}$ and $Z_{11}^L$ respectively.

\subsection{Infrared factorization}

The resulting UV finite amplitudes will contain divergences of infrared origin, which
remain as poles in the 
dimensional regularization parameter $\epsilon$.  These will cancel 
when combined with the real emission processes to compute observables.
While these divergences disappear in the physical observables, the amplitudes beyond leading order 
demonstrate a very rich universal structure in the IR region. 
Catani~\cite{Catani:1998bh} predicted IR divergences for $n$-point two-loop amplitudes in terms of certain
universal IR anomalous dimensions, exploiting the iterative structure of the IR singular parts 
in any UV renormalized amplitudes in QCD. These could be related \cite{Sterman:2002qn} to the factorization and resummation properties of QCD amplitudes, and were subsequently generalized to higher loop 
order~\cite{Becher:2009cu,Gardi:2009qi}. 
Following \cite{Catani:1998bh}, we obtain
\begin{eqnarray}
\label{mfin}
\mathcal{M}_{i} ^{B,(0)} &=& \mathcal{M}_{i} ^{B,(0)}
\nonumber\\
\mathcal{M}_{i} ^{B,(1)} &=& 2\mathbf{I}_{g}^{(1)}(\epsilon)\mathcal{M}_{i} ^{B,(0)} + \mathcal{M}_{i} ^{B,(1),fin}
\nonumber\\
\mathcal{M}_{i} ^{B,(2)} &=& 4\mathbf{I}_{g}^{(2)}(\epsilon)\mathcal{M}_{i} ^{B,(0)} + 2\mathbf{I}_{g}^{(1)}(\epsilon)\mathcal{M}_{i} ^{B,(1)} + \mathcal{M}_{i} ^{B,(2),fin}
\end{eqnarray}
where $\mathbf{I}_{g}^{(1)}(\epsilon), \,\mathbf{I}_{g}^{(2)}(\epsilon) $ are the IR singularity operators given by
\begin{eqnarray}
\label{isub}
\!\!\!\!\!\!\!\!\!\!\!\mathbf{I}_{g}^{(1)}(\epsilon) &=&  -\frac{e^{\epsilon \gamma }}{\Gamma \left( 1 - \epsilon\right)}
\left(  \frac{ C_A}{\epsilon^2} + \frac{\beta_0}{2\epsilon} \right) \left( -\frac{\mu_R^2}{s}\right)^{\epsilon} ,
\nonumber\\
\!\!\!\!\!\!\!\!\!\!\! \mathbf{I}_{g}^{(2)}(\epsilon) &=& 
-\frac{1}{2}\mathbf{I}^{(1)}_g(\epsilon)\left[ \mathbf{I}^{(1)}_g(\epsilon) + 
\frac{\beta_0}{\epsilon} \right] + 
\frac{e^{-\epsilon\gamma}\Gamma(1-2\epsilon)}{\Gamma(1-\epsilon)}\left[\frac{\beta_0}{2\epsilon}+K\right]\mathbf{I}_{g}^{(1)}(2\epsilon)+
2\mathbf{H}^{(2)}_g(\epsilon),
\end{eqnarray}
with
 \begin{align}
 K &=\left( \frac{67}{18}-\zeta_2\right)C_A - \frac{10}{9}T_F n_f,
 \nonumber\\
 \mathbf{H}^{(2)}_{g}(\epsilon) &= -\left(-\frac{\mu_{R}^2}{s}\right)^{2\epsilon} \frac{e^{\epsilon\gamma}}{\Gamma(1-\epsilon)} 
\nonumber\\
&\times \frac{1}{2\epsilon}\left\{ 
C_A^2\left(-\frac{5}{24}-\frac{11}{48}\zeta_2-\frac{1}{4}\zeta_3\right) + C_A 
n_f\left(\frac{29}{54} + \frac{1}{24}\zeta_2\right)-\frac{1}{4}C_F n_f 
-\frac{5}{54}n_f^2\right\}.
\end{align}
It is known that the terms that become finite or vanish as $\epsilon$ goes to zero, 
i.e., ${\cal O}(\epsilon^\alpha), \alpha \ge 0$  in the subtraction operators $\mathbf{I}^{(1)}_g$ and $\mathbf{I}^{(2)}_g$ are 
arbitrary and they define the scheme in which these IR divergences are subtracted to obtain IR 
finite parts of amplitudes,  ${\cal M}^{B,(j),fin}_i$.
These scheme-dependent terms in the finite part of virtual contributions will 
cancel against those coming from the soft gluon emission subprocesses at the observable level. 
The only scheme dependence that will be left
in a physical subprocess coefficient function is then due to the subtraction of collinear initial state divergences
through mass factorization, 
parametrized by a factorization scale $\mu_F$.


\section{Calculation of the Amplitude}
\label{sec:calc}
For the amplitudes of class-B, we needed to consider only those diagrams which involve a pair of vertices resulting
from the first term of the effective Lagrangian and hence all the amplitudes are proportional to $C_H^2$.
These Feynman diagrams up to two-loop level were obtained with help of the  package 
QGRAF \cite{Nogueira:1991ex}. There are 2 diagrams
at tree level, 37 at one loop  and 865 at two-loop order in perturbation theory. 
The output from QGRAF was then used for further algebraic manipulations involving  traces of Dirac matrices, contraction 
of Lorentz and color indices, using 
two independent sets of in-house routines based on a symbolic package FORM \cite{Vermaseren:2000nd}. The entire
manipulations were performed in $d = 4 -2 \epsilon$ dimensions and most of the algebraic simplifications were done
at this stage.  We used the Feynman gauge throughout and hence allowed ghost particles in the
loops. External ghosts are not required due to the transversal nature of the tensorial projectors Eq.~(\ref{eq:proj}). 

At this stage, we obtain a large number of Feynman integrals with different sets of propagators and each
containing scalar products of the independent external and internal momenta. 
Using the REDUZE2 package~\cite{vonManteuffel:2012np}, we can identify the momentum shifts that are 
required to express each diagram in terms of a standard set of propagators (called auxiliary topology). The 
auxiliary topologies in 
the two-loop corrections to the class-B process are identical to those in equal-mass 
on-shell vector boson pair production at this 
loop order. They are described in~\cite{Gehrmann:2014bfa} and were used to compute the 
two-loop corrections to $q\bar q\to VV$ in~\cite{Cascioli:2014yka,Gehrmann:2014fva}. They were subsequently extended 
towards non-equal gauge boson masses~\cite{Caola:2014iua,Gehrmann:2015ora,vonManteuffel:2015msa,Caola:2015ila}.

It is well known that the resulting Feynman integrals are not all independent and hence they can be expressed
in terms of fewer scalar integrals, called Master Integrals (MIs) by using
integration-by-parts (IBP) identities~\cite{Tkachov:1981wb,Chetyrkin:1981qh}. 
 Further simplifications can be
done by exploiting the Lorentz invariance of the integrands, resulting in Lorentz invariance (LI) 
identities~\cite{Gehrmann:1999as}.  These identities can be solved systematically using 
lexicographic ordering (Laporta algorithm,~\cite{Laporta:2001dd}) to 
express any Feynman integral in terms of master integrals. 
These are implemented in several specialized computer algebra packages, for example
AIR \cite{Anastasiou:2004vj}, FIRE \cite{Smirnov:2008iw},  
REDUZE2 \cite{Studerus:2009ye,vonManteuffel:2012np} and LiteRed \cite{Lee:2013mka},  
to perform suitable integral reductions such that one ends up with only MIs.  
We performed two independent reductions of the integrals in the two-loop 
class-B amplitude, one based on 
 the Mathematica based package LiteRed \cite{Lee:2013mka} and the other based on REDUZE2~\cite{vonManteuffel:2012np}.
Counting kinematical crossings as independent integrals, we can express the one-loop amplitude in terms of 
10 master integrals, while the two-loop amplitude contains 149 master integrals. 
These master integrals are 
two-loop four-point functions with internal massless propagators and two massive external legs of equal mass. They 
were computed analytically as Laurent series  expansion  in $\epsilon$  in~\cite{Gehrmann:2013cxs,Gehrmann:2014bfa}. 

These MIs were then expressed in terms of generalized harmonic polylogarithms. 
An alternative functional basis can be obtained in terms of logarithms, polylogarithms $\rm{Li}_{n\leq 4}$ and the 
multiple polylogarithm $\rm{Li}_{2,2}$ by matching the original expression at the symbol level~\cite{Gehrmann:2014bfa}. 
We use the master integrals in this latter representation. 
Substituting the MIs from \cite{Gehrmann:2013cxs,Gehrmann:2014bfa}, 
we obtain the bare amplitudes ${\cal \hat{M}}_{i}^{B,(1)}$ and  ${\cal \hat{M}}_{i}^{B,(2)}$. 
The ultraviolet singularities present in these amplitudes are removed by 
renormalisation as described in Section~\ref{sec:ren} above. 
The resulting UV renormalized amplitudes contain only infrared  divergences. 
We find that the poles of these amplitudes agree with what is expected from IR factorization, Eq.~(\ref{mfin}),
using the subtraction operators of Eq.~(\ref{isub}). These define 
the finite remainders of the amplitudes $\mathcal{M}_i^{B,(j), fin}$ with $j=0,1,2$.  


\section{Numerical Evaluation of the Two-loop Amplitudes}
\label{sec:num}
\begin{figure}[t]
\centerline{
 \includegraphics[width=8cm,height=9cm,angle=0]{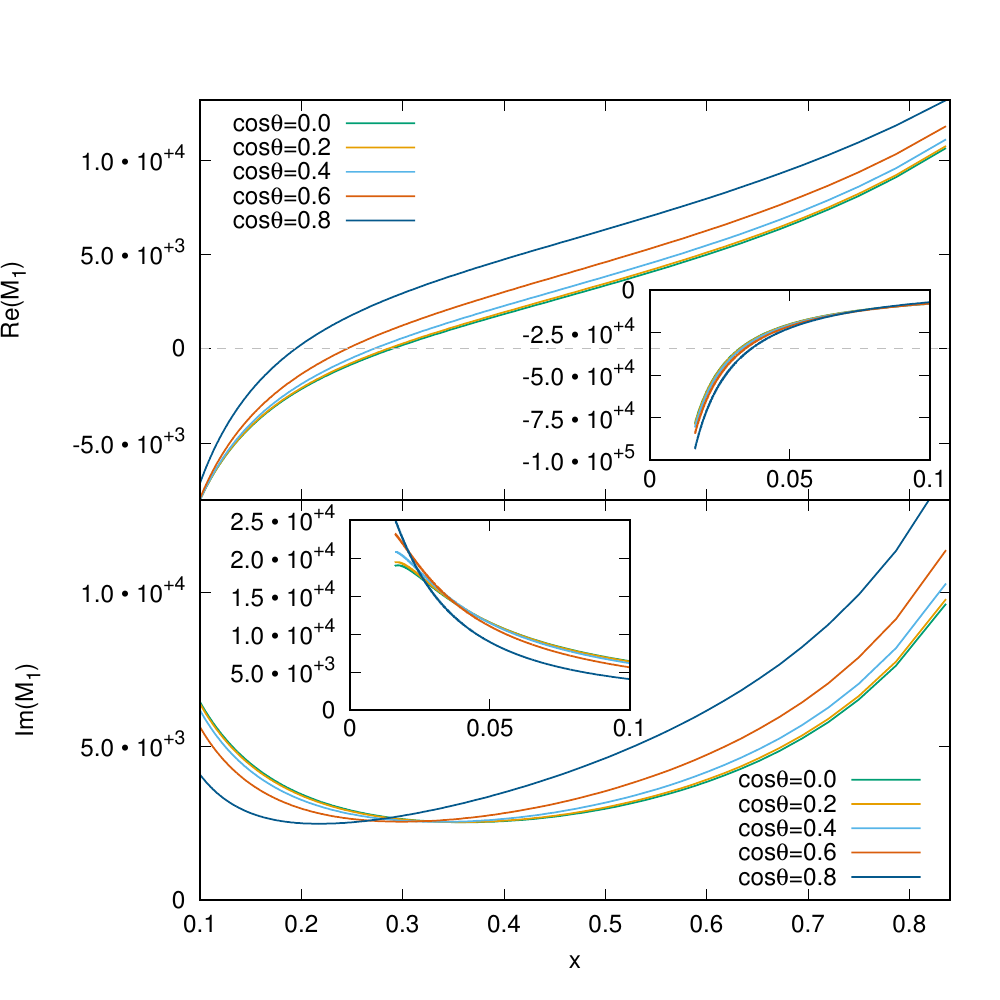}
  \includegraphics[width=8cm,height=9cm,angle=0]{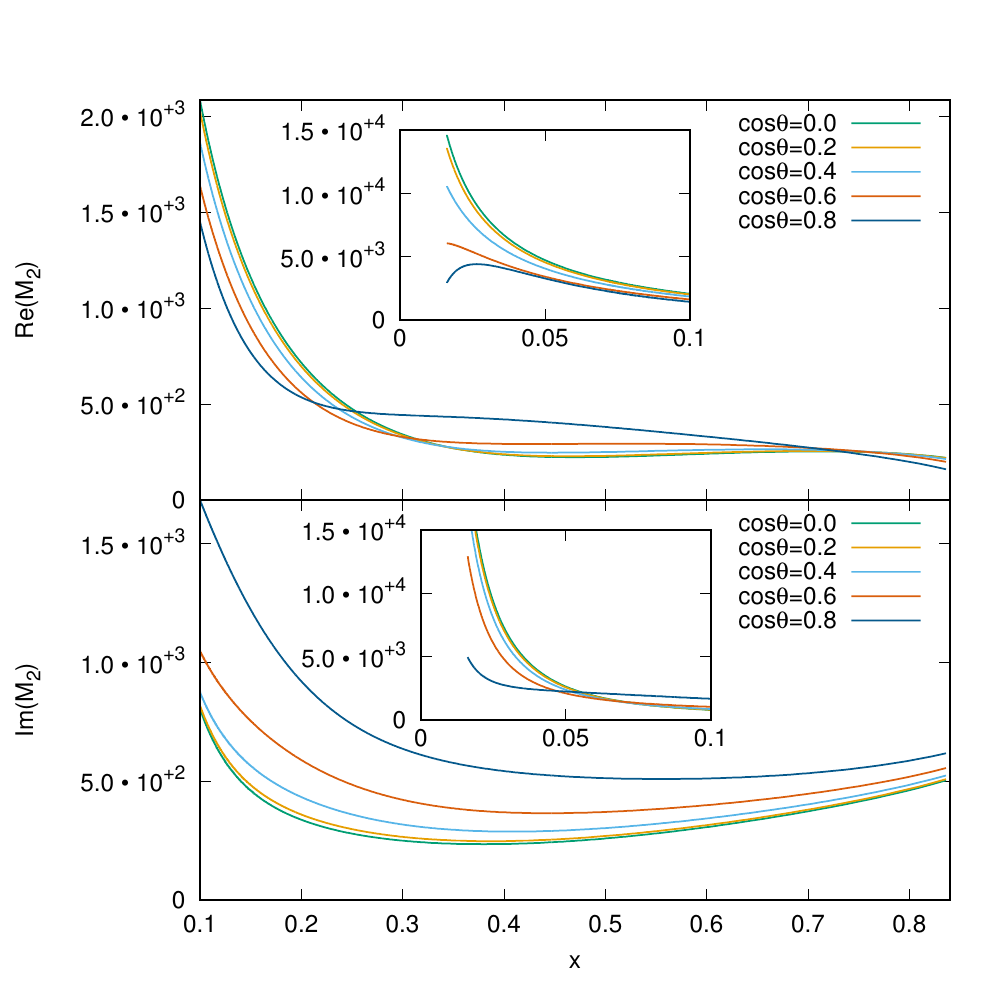}
}
\caption{\sf Behavior of ${\cal M}_1^{B,(2),fin}$  (left) and  ${\cal M}_2^{B,(2),fin}$ (right) 
as a function of the scaling variable $x$
for different values of $\cos(\theta)$. The insets show the region close to $x=0$.}
\label{Amp1}
\end{figure}

The finite remainders ${\cal M}_i^{B,(j),fin}$, $i=1,2$, computed
in the previous section,
are expressed in terms of multiple classical polylogarithms with
arguments depending on the
scaling variables $x,y$ and their coefficients further depending on the
Higgs mass $m_h$.
Since the analytical results are too long to be presented in this
article, we have provided
ancillary files containing the analytical results (and their numerical
evaluation) in Mathematica
format. In order to demonstrate the dependence of the two-loop finite
remainders on the
scaling variables $x$ and $y$ for $m_h=125$ GeV and with   $\mu_R^2=m_h^2/2$,
we plot real and
imaginary parts of
both ${\cal M}_i^{B,(2),fin}$, $i=1,2$ as a function of the partonic invariant mass variable $x$ for
different choices of
$\cos(\theta)$, where $\theta$ is the angle between one of
the Higgs bosons in the final state and one of the
initial gluons in their center of mass frame. We extract an additional factor of 
$m_h^2$ in the plots. The amplitude is invariant under 
$\cos(\theta) \to -\cos(\theta)$, as expected for a purely bosonic amplitude. Since this symmetry has not 
been used  in the setup of the calculation, it serves as a strong check on 
our results.

In Fig.~\ref{Amp1}, we display the real and imaginary parts of the 
amplitude  ${\cal M}_1^{B,(2),fin}$ (left panel) and ${\cal M}_2^{B,(2),fin}$ (right panel). 
  The behavior of the 
amplitudes close to the production threshold, $x=0$, is  shown in the insets. We see that the finite parts of the 
two-loop amplitude shows stable behavior, and they display a non-trivial dependence on the
process kinematics.

In the numerical evaluations, the large rational coefficients of
the classical polylogarithms can introduce numerical instabilities in
case we do not demand high enough precision. In particular, there are
large cancellations between the numerator and denominator of rational
functions. Therefore, we evaluate the polylogarithms at double, and the
rational coefficients at even higher precision.


\section{Discussion and Conclusions}
\label{sec:conc}
The two-loop massless corrections to the $g+g\to H+H$ amplitude derived above complete the set of 
purely virtual amplitudes required for the prediction of the N${}^3$LO  corrections to 
Higgs boson pair production in gluon fusion, in the infinite top quark mass limit. 
All other amplitudes relevant at this order are either (class-A) known already from the calculation of inclusive gluon fusion 
Higgs boson production at N${}^3$LO~\cite{Anastasiou:2015ema,Mistlberger:2018etf} or 
(class-B) amount to one-loop and tree-level amplitudes that can be  
computed
using automated tools. 
The combination of these amplitudes into a fully differential N${}^3$LO calculation of Higgs boson 
pair production does still require substantial advances in the techniques for handling infrared singular real radiation 
configurations at this order, with first steps being taken most recently~\cite{Currie:2018fgr,Cieri:2018oms}.  

More imminent applications of the newly derived results to Higgs boson pair production 
 are the computation of fixed order 
 soft-virtual corrections to the total 
 cross section or of the hard matching coefficients in the  resummation of corrections at low 
 pair transverse momentum.  
 
In this paper, we have computed all virtual amplitudes that contribute 
to the production of a pair of Higgs bosons from the gluon-gluon
initiated partonic processes at order $a_s^4$. The calculation is performed 
in an effective field theory where the top quark is integrated out, and all other quarks are massless. The exact 
calculation of top quark mass effects is currently out of reach at this order, but reweighting procedures allow 
to reliably quantify these effects~\cite{Grazzini:2018bsd}.  
  We deal with two classes of amplitudes
separately, named class-A  (one effective operator insertion) 
and class-B (two effective operator insertions).  The amplitudes of class-A can be related to the gluon 
form factor which is already known up to three loop order while amplitudes of class-B were known previously 
up to one loop. Our explicit computation of the two-loop corrections to the class-B amplitudes 
now completes the perturbative expansion of the $g+g\to H+H$ amplitude to order $a_s^4$. 
We observe that the pole structure of the amplitude is in agreement with predictions from infrared factorization, 
and provide (as ancillary files with the arXiv submission of this article) a numerical code to evaluate its finite 
remainder piece. The newly derived amplitudes open up opportunities for a new level of precision
 phenomenology predictions in Higgs boson pair production.  

\section*{Acknowledgements}
We would like to thank  Claude Duhr, Anirban Karan, Narayan Rana, Lorenzo Tancredi and Andreas von Manteuffel for several useful discussions. 
This research was supported in part by the Swiss National Science Foundation (SNF) under contract 200020-175595,
by the Pauli Center for Theoretical Studies and by the Research Executive Agency (REA) of the European Union under 
the ERC Advanced Grant MC@NNLO (340983) and ERC Starting Grant MathAm (39568).


\begin{thebibliography}{10}

\bibitem{Aad:2012tfa}
{\scshape ATLAS} collaboration, G.~Aad et~al., \emph{{Observation of a new
  particle in the search for the Standard Model Higgs boson with the ATLAS
  detector at the LHC}},
  \href{https://doi.org/10.1016/j.physletb.2012.08.020}{\emph{Phys. Lett.}
  {\bfseries B716} (2012) 1--29},
  [\href{https://arxiv.org/abs/1207.7214}{{\ttfamily 1207.7214}}].

\bibitem{Chatrchyan:2012xdj}
{\scshape CMS} collaboration, S.~Chatrchyan et~al., \emph{{Observation of a new
  boson at a mass of 125 GeV with the CMS experiment at the LHC}},
  \href{https://doi.org/10.1016/j.physletb.2012.08.021}{\emph{Phys. Lett.}
  {\bfseries B716} (2012) 30--61},
  [\href{https://arxiv.org/abs/1207.7235}{{\ttfamily 1207.7235}}].

\bibitem{Englert:2014uua}
C.~Englert, A.~Freitas, M.~M. {M\"uhlleitner}, T.~Plehn, M.~Rauch, M.~Spira and
  K.~Walz, \emph{{Precision Measurements of Higgs Couplings: Implications for
  New Physics Scales}},
  \href{https://doi.org/10.1088/0954-3899/41/11/113001}{\emph{J. Phys.}
  {\bfseries G41} (2014) 113001},
  [\href{https://arxiv.org/abs/1403.7191}{{\ttfamily 1403.7191}}].

\bibitem{Dawson:1998py}
S.~Dawson, S.~Dittmaier and M.~Spira, \emph{{Neutral Higgs boson pair
  production at hadron colliders: QCD corrections}},
  \href{https://doi.org/10.1103/PhysRevD.58.115012}{\emph{Phys. Rev.}
  {\bfseries D58} (1998) 115012},
  [\href{https://arxiv.org/abs/hep-ph/9805244}{{\ttfamily hep-ph/9805244}}].

\bibitem{Djouadi:1999gv}
A.~Djouadi, W.~Kilian, M.~{M\"uhlleitner} and P.~M. Zerwas, \emph{{Testing
  Higgs selfcouplings at $e^+e^-$ linear colliders}},
  \href{https://doi.org/10.1007/s100529900082}{\emph{Eur. Phys. J.} {\bfseries
  C10} (1999) 27--43}, [\href{https://arxiv.org/abs/hep-ph/9903229}{{\ttfamily
  hep-ph/9903229}}].

\bibitem{Djouadi:1999rca}
A.~Djouadi, W.~Kilian, M.~{M\"uhlleitner} and P.~M. Zerwas, \emph{{Production
  of neutral Higgs boson pairs at LHC}},
  \href{https://doi.org/10.1007/s100529900083}{\emph{Eur. Phys. J.} {\bfseries
  C10} (1999) 45--49}, [\href{https://arxiv.org/abs/hep-ph/9904287}{{\ttfamily
  hep-ph/9904287}}].

\bibitem{Muhlleitner:2000jj}
M.~M. {M\"uhlleitner}, \emph{{Higgs particles in the standard model and
  supersymmetric theories}}, Ph.D. thesis, Hamburg U., 2000.
\newblock \href{https://arxiv.org/abs/hep-ph/0008127}{{\ttfamily
  hep-ph/0008127}}.

\bibitem{Glover:1987nx}
E.~W.~N. Glover and J.~J. van~der Bij, \emph{{Higgs boson pair production via
  gluon fusion}},
  \href{https://doi.org/10.1016/0550-3213(88)90083-1}{\emph{Nucl. Phys.}
  {\bfseries B309} (1988) 282--294}.

\bibitem{Plehn:1996wb}
T.~Plehn, M.~Spira and P.~M. Zerwas, \emph{{Pair production of neutral Higgs
  particles in gluon-gluon collisions}},
  \href{https://doi.org/10.1016/0550-3213(96)00418-X,
  10.1016/S0550-3213(98)00406-4}{\emph{Nucl. Phys.} {\bfseries B479} (1996)
  46--64}, [\href{https://arxiv.org/abs/hep-ph/9603205}{{\ttfamily
  hep-ph/9603205}}].

\bibitem{Baglio:2012np}
J.~Baglio, A.~Djouadi, R.~{Gr\"ober}, M.~M. {M\"uhlleitner}, J.~Quevillon and
  M.~Spira, \emph{{The measurement of the Higgs self-coupling at the LHC:
  theoretical status}},
  \href{https://doi.org/10.1007/JHEP04(2013)151}{\emph{JHEP} {\bfseries 04}
  (2013) 151}, [\href{https://arxiv.org/abs/1212.5581}{{\ttfamily 1212.5581}}].

\bibitem{Barger:2013jfa}
V.~Barger, L.~L. Everett, C.~B. Jackson and G.~Shaughnessy, \emph{{Higgs-Pair
  Production and Measurement of the Triscalar Coupling at LHC(8,14)}},
  \href{https://doi.org/10.1016/j.physletb.2013.12.013}{\emph{Phys. Lett.}
  {\bfseries B728} (2014) 433--436},
  [\href{https://arxiv.org/abs/1311.2931}{{\ttfamily 1311.2931}}].

\bibitem{Dolan:2012rv}
M.~J. Dolan, C.~Englert and M.~Spannowsky, \emph{{Higgs self-coupling
  measurements at the LHC}},
  \href{https://doi.org/10.1007/JHEP10(2012)112}{\emph{JHEP} {\bfseries 10}
  (2012) 112}, [\href{https://arxiv.org/abs/1206.5001}{{\ttfamily 1206.5001}}].

\bibitem{Papaefstathiou:2012qe}
A.~Papaefstathiou, L.~L. Yang and J.~Zurita, \emph{{Higgs boson pair production
  at the LHC in the $b \bar{b} W^+ W^-$ channel}},
  \href{https://doi.org/10.1103/PhysRevD.87.011301}{\emph{Phys. Rev.}
  {\bfseries D87} (2013) 011301},
  [\href{https://arxiv.org/abs/1209.1489}{{\ttfamily 1209.1489}}].

\bibitem{deLima:2014dta}
D.~E. Ferreira~de Lima, A.~Papaefstathiou and M.~Spannowsky, \emph{{Standard
  model Higgs boson pair production in the ($ b\overline{b} $)($ b\overline{b}
  $) final state}}, \href{https://doi.org/10.1007/JHEP08(2014)030}{\emph{JHEP}
  {\bfseries 08} (2014) 030},
  [\href{https://arxiv.org/abs/1404.7139}{{\ttfamily 1404.7139}}].

\bibitem{Behr:2015oqq}
J.~K. Behr, D.~Bortoletto, J.~A. Frost, N.~P. Hartland, C.~Issever and J.~Rojo,
  \emph{{Boosting Higgs pair production in the $b\bar{b}b\bar{b}$ final state
  with multivariate techniques}},
  \href{https://doi.org/10.1140/epjc/s10052-016-4215-5}{\emph{Eur. Phys. J.}
  {\bfseries C76} (2016) 386},
  [\href{https://arxiv.org/abs/1512.08928}{{\ttfamily 1512.08928}}].

\bibitem{Grober:2017gut}
R.~{Gr\"ober}, M.~{M\"uhlleitner} and M.~Spira, \emph{{Higgs Pair Production at
  NLO QCD for CP-violating Higgs Sectors}},
  \href{https://doi.org/10.1016/j.nuclphysb.2017.10.002}{\emph{Nucl. Phys.}
  {\bfseries B925} (2017) 1--27},
  [\href{https://arxiv.org/abs/1705.05314}{{\ttfamily 1705.05314}}].

\bibitem{Grigo:2013rya}
J.~Grigo, J.~Hoff, K.~Melnikov and M.~Steinhauser, \emph{{On the Higgs boson
  pair production at the LHC}},
  \href{https://doi.org/10.1016/j.nuclphysb.2013.06.024}{\emph{Nucl. Phys.}
  {\bfseries B875} (2013) 1--17},
  [\href{https://arxiv.org/abs/1305.7340}{{\ttfamily 1305.7340}}].

\bibitem{Frederix:2014hta}
R.~Frederix, S.~Frixione, V.~Hirschi, F.~Maltoni, O.~Mattelaer, P.~Torrielli,
  E.~Vryonidou and M.~Zaro, \emph{{Higgs pair production at the LHC with NLO
  and parton-shower effects}},
  \href{https://doi.org/10.1016/j.physletb.2014.03.026}{\emph{Phys. Lett.}
  {\bfseries B732} (2014) 142--149},
  [\href{https://arxiv.org/abs/1401.7340}{{\ttfamily 1401.7340}}].

\bibitem{Maltoni:2014eza}
F.~Maltoni, E.~Vryonidou and M.~Zaro, \emph{{Top-quark mass effects in double
  and triple Higgs production in gluon-gluon fusion at NLO}},
  \href{https://doi.org/10.1007/JHEP11(2014)079}{\emph{JHEP} {\bfseries 11}
  (2014) 079}, [\href{https://arxiv.org/abs/1408.6542}{{\ttfamily 1408.6542}}].

\bibitem{Degrassi:2016vss}
G.~Degrassi, P.~P. Giardino and R.~{Gr\"ober}, \emph{{On the two-loop virtual
  QCD corrections to Higgs boson pair production in the Standard Model}},
  \href{https://doi.org/10.1140/epjc/s10052-016-4256-9}{\emph{Eur. Phys. J.}
  {\bfseries C76} (2016) 411},
  [\href{https://arxiv.org/abs/1603.00385}{{\ttfamily 1603.00385}}].


\bibitem{Grober:2017uho}
R.~{Gr\"ober}, A.~Maier and T.~Rauh, \emph{{Reconstruction of top-quark mass
  effects in Higgs pair production and other gluon-fusion processes}},
  \href{https://doi.org/10.1007/JHEP03(2018)020}{\emph{JHEP} {\bfseries 03}
  (2018) 020}, [\href{https://arxiv.org/abs/1709.07799}{{\ttfamily
  1709.07799}}].

\bibitem{Bonciani:2018omm}
R.~Bonciani, G.~Degrassi, P.~P. Giardino and R.~{Gr\"ober}, \emph{{Analytical
  Method for Next-to-Leading-Order QCD Corrections to Double-Higgs
  Production}},
  \href{https://doi.org/10.1103/PhysRevLett.121.162003}{\emph{Phys. Rev. Lett.}
  {\bfseries 121} (2018) 162003},
  [\href{https://arxiv.org/abs/1806.11564}{{\ttfamily 1806.11564}}].

\bibitem{Borowka:2016ehy}
S.~Borowka, N.~Greiner, G.~Heinrich, S.~Jones, M.~Kerner, J.~Schlenk,
  U.~Schubert and T.~Zirke, \emph{{Higgs Boson Pair Production in Gluon Fusion
  at Next-to-Leading Order with Full Top-Quark Mass Dependence}},
  \href{https://doi.org/10.1103/PhysRevLett.117.079901,
  10.1103/PhysRevLett.117.012001}{\emph{Phys. Rev. Lett.} {\bfseries 117}
  (2016) 012001}, [\href{https://arxiv.org/abs/1604.06447}{{\ttfamily
  1604.06447}}].

\bibitem{Borowka:2016ypz}
S.~Borowka, N.~Greiner, G.~Heinrich, S.~P. Jones, M.~Kerner, J.~Schlenk and
  T.~Zirke, \emph{{Full top quark mass dependence in Higgs boson pair
  production at NLO}},
  \href{https://doi.org/10.1007/JHEP10(2016)107}{\emph{JHEP} {\bfseries 10}
  (2016) 107}, [\href{https://arxiv.org/abs/1608.04798}{{\ttfamily
  1608.04798}}].

\bibitem{deFlorian:2013uza}
D.~de~Florian and J.~Mazzitelli, \emph{{Two-loop virtual corrections to Higgs
  pair production}},
  \href{https://doi.org/10.1016/j.physletb.2013.06.046}{\emph{Phys. Lett.}
  {\bfseries B724} (2013) 306--309},
  [\href{https://arxiv.org/abs/1305.5206}{{\ttfamily 1305.5206}}].

\bibitem{Grigo:2015dia}
J.~Grigo, J.~Hoff and M.~Steinhauser, \emph{{Higgs boson pair production: top
  quark mass effects at NLO and NNLO}},
  \href{https://doi.org/10.1016/j.nuclphysb.2015.09.012}{\emph{Nucl. Phys.}
  {\bfseries B900} (2015) 412--430},
  [\href{https://arxiv.org/abs/1508.00909}{{\ttfamily 1508.00909}}].

\bibitem{deFlorian:2013jea}
D.~de~Florian and J.~Mazzitelli, \emph{{Higgs Boson Pair Production at
  Next-to-Next-to-Leading Order in QCD}},
  \href{https://doi.org/10.1103/PhysRevLett.111.201801}{\emph{Phys. Rev. Lett.}
  {\bfseries 111} (2013) 201801},

\bibitem{Grigo:2014jma}
J.~Grigo, K.~Melnikov and M.~Steinhauser, \emph{{Virtual corrections to Higgs
  boson pair production in the large top quark mass limit}},
  \href{https://doi.org/10.1016/j.nuclphysb.2014.09.003}{\emph{Nucl. Phys.}
  {\bfseries B888} (2014) 17--29},
  [\href{https://arxiv.org/abs/1408.2422}{{\ttfamily 1408.2422}}].

\bibitem{Li:2013flc}
Q.~Li, Q.-S. Yan and X.~Zhao, \emph{{Higgs Pair Production: Improved
  Description by Matrix Element Matching}},
  \href{https://doi.org/10.1103/PhysRevD.89.033015}{\emph{Phys. Rev.}
  {\bfseries D89} (2014) 033015},
  [\href{https://arxiv.org/abs/1312.3830}{{\ttfamily 1312.3830}}].

\bibitem{Maierhofer:2013sha}
P.~{Maierh\"ofer} and A.~Papaefstathiou, \emph{{Higgs Boson pair production
  merged to one jet}},
  \href{https://doi.org/10.1007/JHEP03(2014)126}{\emph{JHEP} {\bfseries 03}
  (2014) 126}, [\href{https://arxiv.org/abs/1401.0007}{{\ttfamily 1401.0007}}].

\bibitem{deFlorian:2015moa}
D.~de~Florian and J.~Mazzitelli, \emph{{Higgs pair production at
  next-to-next-to-leading logarithmic accuracy at the LHC}},
  \href{https://doi.org/10.1007/JHEP09(2015)053}{\emph{JHEP} {\bfseries 09}
  (2015) 053}, [\href{https://arxiv.org/abs/1505.07122}{{\ttfamily
  1505.07122}}].

\bibitem{Grazzini:2018bsd}
M.~Grazzini, G.~Heinrich, S.~Jones, S.~Kallweit, M.~Kerner, J.~M. Lindert and
  J.~Mazzitelli, \emph{{Higgs boson pair production at NNLO with top quark mass
  effects}}, \href{https://doi.org/10.1007/JHEP05(2018)059}{\emph{JHEP}
  {\bfseries 05} (2018) 059},
  [\href{https://arxiv.org/abs/1803.02463}{{\ttfamily 1803.02463}}].

\bibitem{Shao:2013bz}
D.~Y. Shao, C.~S. Li, H.~T. Li and J.~Wang, \emph{{Threshold resummation
  effects in Higgs boson pair production at the LHC}},
  \href{https://doi.org/10.1007/JHEP07(2013)169}{\emph{JHEP} {\bfseries 07}
  (2013) 169}, [\href{https://arxiv.org/abs/1301.1245}{{\ttfamily 1301.1245}}].

\bibitem{Djouadi:1991tka}
A.~Djouadi, M.~Spira and P.~M. Zerwas, \emph{{Production of Higgs bosons in
  proton colliders: QCD corrections}},
  \href{https://doi.org/10.1016/0370-2693(91)90375-Z}{\emph{Phys. Lett.}
  {\bfseries B264} (1991) 440--446}.

\bibitem{Kramer:1996iq}
M.~Kramer, E.~Laenen and M.~Spira, \emph{{Soft gluon radiation in Higgs boson
  production at the LHC}},
  \href{https://doi.org/10.1016/S0550-3213(97)00679-2}{\emph{Nucl. Phys.}
  {\bfseries B511} (1998) 523--549},
  [\href{https://arxiv.org/abs/hep-ph/9611272}{{\ttfamily hep-ph/9611272}}].

\bibitem{Chetyrkin:1997iv}
K.~G. Chetyrkin, B.~A. Kniehl and M.~Steinhauser, \emph{{Hadronic Higgs decay
  to order $\alpha_s^4$}},
  \href{https://doi.org/10.1103/PhysRevLett.79.353}{\emph{Phys. Rev. Lett.}
  {\bfseries 79} (1997) 353--356},
  [\href{https://arxiv.org/abs/hep-ph/9705240}{{\ttfamily hep-ph/9705240}}].

\bibitem{Spira:2016zna}
M.~Spira, \emph{{Effective Multi-Higgs Couplings to Gluons}},
  \href{https://doi.org/10.1007/JHEP10(2016)026}{\emph{JHEP} {\bfseries 10}
  (2016) 026}, [\href{https://arxiv.org/abs/1607.05548}{{\ttfamily
  1607.05548}}].

\bibitem{Gerlach:2018hen}
M.~Gerlach, F.~Herren and M.~Steinhauser, \emph{{Wilson coefficients for Higgs
  boson production and decoupling relations to
  $\mathcal{O}\left(\alpha_s^4\right)$}},
  \href{https://arxiv.org/abs/1809.06787}{{\ttfamily 1809.06787}}.

\bibitem{Baikov:2009bg}
P.~A. Baikov, K.~G. Chetyrkin, A.~V. Smirnov, V.~A. Smirnov and M.~Steinhauser,
  \emph{{Quark and gluon form factors to three loops}},
  \href{https://doi.org/10.1103/PhysRevLett.102.212002}{\emph{Phys. Rev. Lett.}
  {\bfseries 102} (2009) 212002},
  [\href{https://arxiv.org/abs/0902.3519}{{\ttfamily 0902.3519}}].

\bibitem{Gehrmann:2010ue}
T.~Gehrmann, E.~W.~N. Glover, T.~Huber, N.~Ikizlerli and C.~Studerus,
  \emph{{Calculation of the quark and gluon form factors to three loops in
  QCD}}, \href{https://doi.org/10.1007/JHEP06(2010)094}{\emph{JHEP} {\bfseries
  06} (2010) 094}, [\href{https://arxiv.org/abs/1004.3653}{{\ttfamily
  1004.3653}}].

\bibitem{Zoller:2016iam}
M.~F. Zoller, \emph{{On the renormalization of operator products: the scalar
  gluonic case}}, \href{https://doi.org/10.1007/JHEP04(2016)165}{\emph{JHEP}
  {\bfseries 04} (2016) 165},
  [\href{https://arxiv.org/abs/1601.08094}{{\ttfamily 1601.08094}}].

\bibitem{Nielsen:1975ph}
N.~K. Nielsen, \emph{{Gauge Invariance and Broken Conformal Symmetry}},
  \href{https://doi.org/10.1016/0550-3213(75)90378-8}{\emph{Nucl. Phys.}
  {\bfseries B97} (1975) 527--540}.

\bibitem{Spiridonov:1988md}
V.~P. Spiridonov and K.~G. Chetyrkin, \emph{{Nonleading mass corrections and
  renormalization of the operators $m \bar{\psi} \psi$ and $G^2_{\mu\nu}$}},
  {\emph{Sov. J. Nucl. Phys.} {\bfseries 47} (1988) 522--527}.

\bibitem{Kataev:1981gr}
A.~L. Kataev, N.~V. Krasnikov and A.~A. Pivovarov, \emph{{Two Loop Calculations
  for the Propagators of Gluonic Currents}},
  \href{https://doi.org/10.1016/0550-3213(82)90338-8,
  10.1016/S0550-3213(97)00101-6}{\emph{Nucl. Phys.} {\bfseries B198} (1982)
  508--518}, [\href{https://arxiv.org/abs/hep-ph/9612326}{{\ttfamily
  hep-ph/9612326}}].

\bibitem{Catani:1998bh}
S.~Catani, \emph{{The Singular behavior of QCD amplitudes at two loop order}},
  \href{https://doi.org/10.1016/S0370-2693(98)00332-3}{\emph{Phys. Lett.}
  {\bfseries B427} (1998) 161--171},
  [\href{https://arxiv.org/abs/hep-ph/9802439}{{\ttfamily hep-ph/9802439}}].

\bibitem{Sterman:2002qn}
G.~F. Sterman and M.~E. Tejeda-Yeomans, \emph{{Multiloop amplitudes and
  resummation}},
  \href{https://doi.org/10.1016/S0370-2693(02)03100-3}{\emph{Phys. Lett.}
  {\bfseries B552} (2003) 48--56},
  [\href{https://arxiv.org/abs/hep-ph/0210130}{{\ttfamily hep-ph/0210130}}].

\bibitem{Becher:2009cu}
T.~Becher and M.~Neubert, \emph{{Infrared singularities of scattering
  amplitudes in perturbative QCD}},
  \href{https://doi.org/10.1103/PhysRevLett.102.162001,
  10.1103/PhysRevLett.111.199905}{\emph{Phys. Rev. Lett.} {\bfseries 102}
  (2009) 162001}, [\href{https://arxiv.org/abs/0901.0722}{{\ttfamily
  0901.0722}}].

\bibitem{Gardi:2009qi}
E.~Gardi and L.~Magnea, \emph{{Factorization constraints for soft anomalous
  dimensions in QCD scattering amplitudes}},
  \href{https://doi.org/10.1088/1126-6708/2009/03/079}{\emph{JHEP} {\bfseries
  03} (2009) 079}, [\href{https://arxiv.org/abs/0901.1091}{{\ttfamily
  0901.1091}}].

\bibitem{Nogueira:1991ex}
P.~Nogueira, \emph{{Automatic Feynman graph generation}},
  \href{https://doi.org/10.1006/jcph.1993.1074}{\emph{J. Comput. Phys.}
  {\bfseries 105} (1993) 279--289}.

\bibitem{Vermaseren:2000nd}
J.~A.~M. Vermaseren, \emph{{New features of FORM}},
  \href{https://arxiv.org/abs/math-ph/0010025}{{\ttfamily math-ph/0010025}}.

\bibitem{vonManteuffel:2012np}
A.~von Manteuffel and C.~Studerus, \emph{{Reduze 2 - Distributed Feynman
  Integral Reduction}},  \href{https://arxiv.org/abs/1201.4330}{{\ttfamily
  1201.4330}}.

\bibitem{Gehrmann:2014bfa}
T.~Gehrmann, A.~von Manteuffel, L.~Tancredi and E.~Weihs, \emph{{The two-loop
  master integrals for $q\overline{q} \to VV$}},
  \href{https://doi.org/10.1007/JHEP06(2014)032}{\emph{JHEP} {\bfseries 06}
  (2014) 032}, [\href{https://arxiv.org/abs/1404.4853}{{\ttfamily 1404.4853}}].

\bibitem{Cascioli:2014yka}
F.~Cascioli, T.~Gehrmann, M.~Grazzini, S.~Kallweit, P.~{Maierh\"ofer}, A.~von
  Manteuffel, S.~Pozzorini, D.~Rathlev, L.~Tancredi and E.~Weihs, \emph{{ZZ
  production at hadron colliders in NNLO QCD}},
  \href{https://doi.org/10.1016/j.physletb.2014.06.056}{\emph{Phys. Lett.}
  {\bfseries B735} (2014) 311--313},
  [\href{https://arxiv.org/abs/1405.2219}{{\ttfamily 1405.2219}}].

\bibitem{Gehrmann:2014fva}
T.~Gehrmann, M.~Grazzini, S.~Kallweit, P.~{Maierh\"ofer}, A.~von Manteuffel,
  S.~Pozzorini, D.~Rathlev and L.~Tancredi, \emph{{$W^+W^-$ Production at
  Hadron Colliders in Next to Next to Leading Order QCD}},
  \href{https://doi.org/10.1103/PhysRevLett.113.212001}{\emph{Phys. Rev. Lett.}
  {\bfseries 113} (2014) 212001},
  [\href{https://arxiv.org/abs/1408.5243}{{\ttfamily 1408.5243}}].

\bibitem{Caola:2014iua}
F.~Caola, J.~M. Henn, K.~Melnikov, A.~V. Smirnov and V.~A. Smirnov,
  \emph{{Two-loop helicity amplitudes for the production of two off-shell
  electroweak bosons in quark-antiquark collisions}},
  \href{https://doi.org/10.1007/JHEP11(2014)041}{\emph{JHEP} {\bfseries 11}
  (2014) 041}, [\href{https://arxiv.org/abs/1408.6409}{{\ttfamily 1408.6409}}].

\bibitem{Gehrmann:2015ora}
T.~Gehrmann, A.~von Manteuffel and L.~Tancredi, \emph{{The two-loop helicity
  amplitudes for $ q\overline{q}^{\prime}\to {V}_1{V}_2\to 4 $ leptons}},
  \href{https://doi.org/10.1007/JHEP09(2015)128}{\emph{JHEP} {\bfseries 09}
  (2015) 128}, [\href{https://arxiv.org/abs/1503.04812}{{\ttfamily
  1503.04812}}].

\bibitem{vonManteuffel:2015msa}
A.~von Manteuffel and L.~Tancredi, \emph{{The two-loop helicity amplitudes for
  $gg \to V_1 V_2 \to 4~\mathrm{leptons}$}},
  \href{https://doi.org/10.1007/JHEP06(2015)197}{\emph{JHEP} {\bfseries 06}
  (2015) 197}, [\href{https://arxiv.org/abs/1503.08835}{{\ttfamily
  1503.08835}}].

\bibitem{Caola:2015ila}
F.~Caola, J.~M. Henn, K.~Melnikov, A.~V. Smirnov and V.~A. Smirnov,
  \emph{{Two-loop helicity amplitudes for the production of two off-shell
  electroweak bosons in gluon fusion}},
  \href{https://doi.org/10.1007/JHEP06(2015)129}{\emph{JHEP} {\bfseries 06}
  (2015) 129}, [\href{https://arxiv.org/abs/1503.08759}{{\ttfamily
  1503.08759}}].

\bibitem{Tkachov:1981wb}
F.~V. Tkachov, \emph{{A Theorem on Analytical Calculability of Four Loop
  Renormalization Group Functions}},
  \href{https://doi.org/10.1016/0370-2693(81)90288-4}{\emph{Phys. Lett.}
  {\bfseries 100B} (1981) 65--68}.

\bibitem{Chetyrkin:1981qh}
K.~G. Chetyrkin and F.~V. Tkachov, \emph{{Integration by Parts: The Algorithm
  to Calculate beta Functions in 4 Loops}},
  \href{https://doi.org/10.1016/0550-3213(81)90199-1}{\emph{Nucl. Phys.}
  {\bfseries B192} (1981) 159--204}.

\bibitem{Gehrmann:1999as}
T.~Gehrmann and E.~Remiddi, \emph{{Differential equations for two loop four
  point functions}},
  \href{https://doi.org/10.1016/S0550-3213(00)00223-6}{\emph{Nucl. Phys.}
  {\bfseries B580} (2000) 485--518},
  [\href{https://arxiv.org/abs/hep-ph/9912329}{{\ttfamily hep-ph/9912329}}].

\bibitem{Laporta:2001dd}
S.~Laporta, \emph{{High precision calculation of multiloop Feynman integrals by
  difference equations}}, \href{https://doi.org/10.1016/S0217-751X(00)00215-7,
  10.1142/S0217751X00002157}{\emph{Int. J. Mod. Phys.} {\bfseries A15} (2000)
  5087--5159}, [\href{https://arxiv.org/abs/hep-ph/0102033}{{\ttfamily
  hep-ph/0102033}}].

\bibitem{Anastasiou:2004vj}
C.~Anastasiou and A.~Lazopoulos, \emph{{Automatic integral reduction for higher
  order perturbative calculations}},
  \href{https://doi.org/10.1088/1126-6708/2004/07/046}{\emph{JHEP} {\bfseries
  07} (2004) 046}, [\href{https://arxiv.org/abs/hep-ph/0404258}{{\ttfamily
  hep-ph/0404258}}].

\bibitem{Smirnov:2008iw}
A.~V. Smirnov, \emph{{Algorithm FIRE -- Feynman Integral REduction}},
  \href{https://doi.org/10.1088/1126-6708/2008/10/107}{\emph{JHEP} {\bfseries
  10} (2008) 107}, [\href{https://arxiv.org/abs/0807.3243}{{\ttfamily
  0807.3243}}].

\bibitem{Studerus:2009ye}
C.~Studerus, \emph{{Reduze-Feynman Integral Reduction in C++}},
  \href{https://doi.org/10.1016/j.cpc.2010.03.012}{\emph{Comput. Phys. Commun.}
  {\bfseries 181} (2010) 1293--1300},
  [\href{https://arxiv.org/abs/0912.2546}{{\ttfamily 0912.2546}}].

\bibitem{Lee:2013mka}
R.~N. Lee, \emph{{LiteRed 1.4: a powerful tool for reduction of multiloop
  integrals}}, \href{https://doi.org/10.1088/1742-6596/523/1/012059}{\emph{J.
  Phys. Conf. Ser.} {\bfseries 523} (2014) 012059},
  [\href{https://arxiv.org/abs/1310.1145}{{\ttfamily 1310.1145}}].

\bibitem{Gehrmann:2013cxs}
T.~Gehrmann, L.~Tancredi and E.~Weihs, \emph{{Two-loop master integrals for $q
  \bar{q} \to VV$: the planar topologies}},
  \href{https://doi.org/10.1007/JHEP08(2013)070}{\emph{JHEP} {\bfseries 08}
  (2013) 070}, [\href{https://arxiv.org/abs/1306.6344}{{\ttfamily 1306.6344}}].

\bibitem{Anastasiou:2015ema}
C.~Anastasiou, C.~Duhr, F.~Dulat, F.~Herzog and B.~Mistlberger, \emph{{Higgs
  Boson Gluon-Fusion Production in QCD at Three Loops}},
  \href{https://doi.org/10.1103/PhysRevLett.114.212001}{\emph{Phys. Rev. Lett.}
  {\bfseries 114} (2015) 212001},
  [\href{https://arxiv.org/abs/1503.06056}{{\ttfamily 1503.06056}}].

\bibitem{Mistlberger:2018etf}
B.~Mistlberger, \emph{{Higgs boson production at hadron colliders at N$^{3}$LO
  in QCD}}, \href{https://doi.org/10.1007/JHEP05(2018)028}{\emph{JHEP}
  {\bfseries 05} (2018) 028},
  [\href{https://arxiv.org/abs/1802.00833}{{\ttfamily 1802.00833}}].

\bibitem{Currie:2018fgr}
J.~Currie, T.~Gehrmann, E.~W.~N. Glover, A.~Huss, J.~Niehues and A.~Vogt,
  \emph{{N$^{3}$LO corrections to jet production in deep inelastic scattering
  using the Projection-to-Born method}},
  \href{https://doi.org/10.1007/JHEP05(2018)209}{\emph{JHEP} {\bfseries 05}
  (2018) 209}, [\href{https://arxiv.org/abs/1803.09973}{{\ttfamily
  1803.09973}}].

\bibitem{Cieri:2018oms}
L.~Cieri, X.~Chen, T.~Gehrmann, E.~W.~N. Glover and A.~Huss, \emph{{Higgs boson
  production at the LHC using the $q_T$ subtraction formalism at N$^3$LO QCD}},
  [\href{https://arxiv.org/abs/1807.11501}{{\ttfamily 1807.11501}}].

\end{thebibliography}

\providecommand{\href}[2]{#2}\begingroup\raggedright\endgroup

\bibliographystyle{JHEP}
\end{document}